Subject: quantumprot.tex
Date: Fri, 14 Jul 2000 14:55:22 -0400 (EDT)
From: V N Muthu Kumar <muthu@feynman.princeton.edu>
To: azmy@feynman.princeton.edu

\documentstyle[aps,twocolumn]{revtex}
\begin{document}
\title{ The Charge-Spin Separated Fermi Fluid in the High $T_c$ Cuprates:
 A Quantum Protectorate}
\author{Philip W. Anderson}
\address {Joseph Henry Laboratories of Physics\\
Princeton University, Princeton, NJ 08544}
\maketitle
\begin{abstract}
We find experimental evidence for spin-charge separation in all four
relevant phases of the cuprates. It is argued that this phenomenon
serves to protect the properties of the cuprates from the effects of
impurities and phonons.
\end{abstract}
\section{Introduction}
 Laughlin and Pines\cite{one} have introduced the term ``Quantum
 protectorate'' as a general descriptor of the fact that certain
 states of quantum many-body systems exhibit properties which are
 unaffected by imperfections, impurities and thermal fluctuations.
 They instance the quantum Hall effect, which can be measured to
 $10^{-9}$ accuracy on samples with mean free paths comparable to
 the electron wavelength, and flux quantization in superconductors,
 equivalent to the Josephson frequency relation which again has
 mensuration accuracy and is independent of imperfections and
 scattering. An even simpler example is the rigidity and
 dimensional stability of crystalline solids evinced by the STM.
 Some of these examples exhibit broken symmetry but whether it is
 correct to ascribe broken symmetry to the quantum Hall effect is
 questionable. I would suggest that the source of quantum
 protection is a collective state of the quantum field involved
 such that the individual particles are sufficiently tightly
 coupled that elementary excitations no longer involve a few
 particles but are collective excitations of the whole system,
and therefore, macroscopic behavior is mostly determined by
overall conservation laws.

 The purpose of this paper is, first, to present the overwhelming
 experimental evidence that the metallic states of the high $T_c$
 cuprate superconductors are a quantum protectorate; and second,
 to propose that this particular collective state involves the
 phenomenon of charge-spin separation, and to give indications as
 to why such a state should act like a quantum protectorate.

 \section{Experimental Evidence}
 We may define four regions of the generic phase diagram of the
 cuprates Fig (1): (I) the ``normal'' metallic state near optimal doping,
 widely assumed to be a non-Fermi liquid; (II) The ``spin gap'' or
 pseudogap state, separated from the above by the temperature
 $T^*$, probably a crossover region; (III) The $d$-wave
 superconducting phase; and (IV) the Mott insulating
 antiferromagnet. I shall assume that the ``stripe'' phase when
 encountered is merely an inhomogeneous mixture of (III) and (IV).

 Phase IV, the Mott insulator, is on the face of it charge-spin
 separated. There is a charge gap of $\sim 2 ev$, while the spin
 wave spectrum extends to zero energy. It is understood
 implicitly, but seldom stated, that the spin waves, which are
 Goldstone bosons of the broken symmetry, are weakly scattered by
 phonons and conventional impurities, and not scattered at all in
 the limit $\omega, Q \to 0$: they are in a quantum protectorate,
 because the spin and charge dynamics have become independent, and
 perturbations which interact primarily with charge do not affect
 spin.

 It is the thesis of this paper that phases I, II and III all
 share this property, which is responsible for the anomalies of
 the high $T_c$ cuprates.

 The transport properties of phase I have been particularly well
 studied in YBCO and to a lesser extent in BISCO and ``214''
 ($La-Sr)CuO_4$, and in BISCO particularly very accurate ARPES
 gives us a window on the one-electron spectrum.
 The energy distribution curves have no features indicating phonon
 contributions to the self-energy of the electrons.
 But it is simply
 the scaling of the conductivity as a function of $T$ and
 $\omega$ which gives us the clearest indication.
 $$\sigma=\omega F\Big({T\over \omega}\Big)\ \ .$$
 That is, there is no extraneous energy scale.  In particular,
 phonon scattering would not show the striking linear rise of
 scattering rate ${1\over \tau}\propto \omega$, above the Debye
 frequency, nor would any conventional electron-electron
scattering. The same behavior of the one-electron self-energy is
 shown in ARPES, in marked contrast to conventional metals.
 This behavior is strikingly shown in Fig. 2, from Ref.\cite{Johnson})
 contrasting the phonon dominated self-energy of a $Mo$ surface state with
 that of a cuprate superconductor.
 Both
 observations show that there is little or no effect of phonon
 scattering. (Data are presented in Ref. 4)

 A second peculiar result is the absence of resistivity saturation
 near the Mott limit, which (though not well understood) is seen
 universally in conventional poor metals, and assumed to be
 associated with strong phonon scattering.

 I have elsewhere emphasized the striking observations of
 different relaxation rates for Hall angle and resistivity, which
 have been amply confirmed by measurements of $\theta_H(\omega)$. I
 have shown that this can be explained by charge-spin separation.
 $\tau_H$, too, shows no evidence of phonon or impurity
 scattering.

 The knowledgeable reader may object that $Zn$ and $Ni$
 impurities, which substitute for $Cu$ in the $CuO_2$ planes,
 $\underline{\rm do}$ in fact act as strong scatterers.  These
 impurities act as Kondo scatterers for the spin degrees of freedom:
 they trap a bound spinon, if you like.  But it is easily shown
 that the Kondo effect is enhanced in a spin-charge separated
 system, so that the Kondo temperature may be above room temperature.
 Thus these impurities scatter spinons at the unitarity limit, as
 observed, but do not show magnetism except at high temperature.
 The observation of this strict dichotomy between these two
 scatterers and most others is very good evidence for the quantum
 protectorate and its explanation in terms of spin-charge
 separation. The idea that this dichotomy can be explained by a conventional
 quasiparticle theory\cite{two} is, frankly, not plausible.

 Phase II is the pseudogap state. Here the most striking evidence
 for spin charge separation is the pseudogap itself, which shows up
 as a gap in the one-electron spectrum along the ``anti-nodal''
 directions in $k$ space, while there is no evidence for a gap for
 charge excitations (except in systems with static stripes.)

 Because the phenomena are much complicated by the pseudogap, it
 is not possible to completely eliminate the possibility of impurity
 or phonon scattering, but there is certainly no evidence for
 either.  Much attention has been given to the
 mysterious nature
 of the pseudogap; for instance it has been realized that the
 violation of the Luttinger theorem on the Fermi surface
 essentially excludes conventional renormalized Fermi liquid
 theory in this region.

 The superconductor, phase III, shows, surprisingly, the clearest evidence
of
 all for the quantum protectorate. Almost all of the
 superconductors are self-doped, presumably by non-stoichiometry
 at the level of 10--20\%. The doping centers are only one layer
 away from the cuprate in an insulating region, and as
 demonstrated in my book\cite{three} they should scatter quite efficiently.
If
 so they are necessarily pair-breaking for conventional $d$-wave
 superconductors. Surely this level of pair-breaking impurities
 would lower $T_c$ probably to zero. These is no evidence whatever that
 $T_c$ is even affected by purity level or by phonon scattering,
 which will also be pair-breaking for a $d$-wave.  For instance,
 the optimum $T_c$ in YBCO is achieved not in $YBa_2Cu_3O_7$,
 which is almost the only stoichiometric cuprate, but in
 $YBa_2Cu_3O_{6.93}$, with 7\% charged impurities.  In a very
 true sense, the biggest mystery of high $T_c$ superconductivity
 is that $T_c$ is so high! It seems likely that such a $T_c$ can
 $\underline{\rm only}$ appear in a quantum protectorate;
 certainly this is true of a $d$-wave superconductor.

 The absence of pair-breaking effects is confirmed when we examine
 transport properties, especially the thermal conductivity of the
 superconductors in a magnetic field.\cite{four} The field-sensitive
 thermal conductivity for $T$ well below $T_c$ must be carried by
 quasiparticle excitations in the gap nodes. (It is electronic
 because it shows a Hall-like (Righi -- LeDuc) effect and because
 is all cases it is eventually destroyed by fields $H<<H_{c2}$.)

 A number of theorists have shown that the only possible
 interpretation of the data involves true Dirac Fermions at the
 gap nodes with effectively zero mass $(E \propto | k - k_0|)$.
 The node is not smeared out by impurity scattering, to any degree
 that can be measured, as it would have to be in conventional
 $d$-wave superconductors. This, to  me, is the crucial evidence
 for a new kind of quantum protectorate.

 \section{Spin-Charge Separation}

 The hypothesis which has been put forward\cite{three} for some of this
 behavior is charge-spin separation: that the elementary
 excitations in the normal state are not quasiparticles with the
 quantum numbers of electrons but are solitons which are
 fractionalized electrons, one carrying the spin quantum number
 and the other (or others) the charge. In particular, the crucial
 component of this idea is the $\underline{\rm spinon}$, a neutral
 excitation carrying only the spin quantum number of an electron.
 The spinon has a history dating back to work by Des Cloiseaux and
 Fadeev on the excitation spectrum of the Bethe solution of the 1D
 Heisenberg model, and was explicitly demonstrated by E. Lieb and F. Wu
 for the 1D Hubbard model. But aside from a remark by Landau, its
 possible validity as an excitation in higher dimensions dates to
 the RVB theories stimulated by high $T_c$.\cite{five}.

 Spin-charge separation is a very natural phenomenon in
 interacting Fermi systems from a symmetry point of view\cite{six}
 The Fermi liquid has an additional symmetry which is not
 contained in the underlying Hamiltonian, in that the two
 quasiparticles of
 opposite spins are exactly degenerate  and have the same velocity at all
 points of the Fermi surface. This is symmetry $SO(4)$ for the
 conserved currents at each Fermi surface point since we
 have 4 degenerate real Majorana Fermions. But the interaction
 terms do not have full $SO(4)$ symmetry, since they change sign
 for improper rotations, so the true symmetry of the interacting
 Hamiltonian is $SO4 \div Z_2 = SU2 \times SU2$, i.e., charge times
 spin. A finite kinetic energy supplies a  field along the
 $\uparrow$ direction of the charge $SU(2)$ and reduces it to
 $U(1)$, the conventional gauge symmetry of charged particles.

 The reason why conventional Fermi liquid theory works is that $U$
 renormalizes to irrelevance because of the ultraviolet divergence
 of the ladder diagrams in 3 dimensions or higher. The result is
 the ``effective range'' theory which allows us to approximate the
 interaction terms, for forward scattering, by a scattering length
 $a$, which leads only to irrelevant symmetry-breaking terms.  In
 one dimension there is no ultraviolet divergence, this does not
 happen, and spin-charge separation  always occurs. 2 is the
 critical dimension and I have shown that in fact there is always
 a marginally relevant term resulting from $U$, when there is
 spin symmetry. (This argument will be expanded in a forthcoming
 article by PWA and F.D.M. Haldane\cite{six})

 Spin-charge separation tells us that the spectrum of exact
 elementary excitations does not consist of quasiparticles, which
 carry both charge and spin. In the Mott insulator antiferromagnet
 there is a large charge gap and the Goldstone boson excitations
 are spin waves. In the other phases, neither charge nor spin are
 gapped; but nonetheless, the spin spectrum remains distinct and
 reflects the symmetries of the spin system. In particular, in the
 absence of time-reversal breaking, the Kramers degeneracy of the
 electron states reverts to the spin spectrum.

 Unlike the Neel-ordered Mott antiferromagnet, the normal phases
 (I) and (II) are based on a ground state with no broken symmetry,
 presumably a singlet spin liquid. The spin excitations in such a
 fluid are $\underline{\rm spinons}$, spin 1/2, uncharged
 fermion-like objects with linear spectra and finite momenta, in the
 only two cases which have been studied formally. (1D,\cite{seven} and
 relatively weakly interacting 2D\cite{eight}.)
 The latter is our model for the ``normal'' phase (I), a spin
 liquid with a Luttinger Fermi surface at all points of which the
 spinon energy vanishes.

 In Phase II we suppose the spin systems to be in a  state related
 to the  ``$\pi$ flux'' phase of Laughlin,
 equivalent to Affleck and Marston's ``s+id'' RVB. This has been
 extensively studied numerically and with Gutzwiller-projection
 based approximations for the half-filled, insulating case, but not
 in the doped insulator. In a subsequent paper we  will demonstrate
 that the Fermi spin liquid is unstable via  a BCS gap
 formation in the spin sector relative to ``s+id''.

 In both of these two phases the charge spectrum remains ungapped.
 In the ideal, weakly interacting, pure Fermi fluid it consists of
 ``holons'', propagating, particle-like solitons which may have
 charge other than $e$, and anyon statistics. But in the actual
 substance the charge excitations are strongly scattered and
 their low-frequency, long-range dynamics is diffusive.

 A lot of effort has gone into describing this phase and its spin
 excitations using gauge theory (Lee et al, Fisher et al\cite{ten}). These
  groups have indeed found a Fermionic field which appears to
 be equivalent to spinons, so at least in this sense there is
 a third formal treatment of spin-charge separation. These groups,
 too, seem to have less to say abut charge excitations.

 Formal theory for a charge-spin separated superconductor is even
 more rudimentary; the work of Fisher and
 Senthil may
 provide some structure, but their model is missing both impurity
 scattering and long-range Coulomb effects.

 These are actually two sources for the ``quantum protectorate''
 effect, not entirely independent but physically distinct. The
 first is that spinons are relatively weakly scattered because
 they are the ``Goldstone Fermions'' which express fundamental
 symmetries of the spin system. The spinon dynamics in
 low-frequency states is averaged over all configurations of the
 holes, hence effectively is the dynamics of a ``squeezed'',
 smoothed Heisenberg-like model with a number of sites equal to
 the number of electrons. Impurities will lead merely to local
 variations of the effective exchange integrals, which are
 inefficient in scattering long-wavelength, low-frequency spin
 fluctuations.

 A second view is more direct. In the charge-spin separated state,
 the electron is a composite particle whose Green's function in
 space-time is the product of charge and spin factors. The
 resulting Fourier transfer $G(k,\omega)$ is the convolution of
 these and is in fact observed in ARPES measurements to have a
 broad, power-law shape with at best a cusp-like feature at the
 (presumed) spinon frequency. Taking either this argument, or the
 ARPES observations, one sees that the one-electron density of states
 vanishes at $\omega =0$, as a power law.

 $$N(\omega)\propto
 \omega^p. \qquad 0<p<1$$
 In the idealized models $p=2\alpha$, twice the Fermi surface exponent,
 but the observations suggest $p\sim 1/2$. Any perturbation which
 couples to electrons, in particular any time-reversal invariant
 perturbation other than substitution in the copper sites, thus
 renormalizes to zero at low frequencies; it can not cause real
 scattering.

 The above discussion holds for the ``normal'' phases I and II.
 For the superconducting phase III am going to make a rather
 radical proposal. This is that the charge excitations essentially
 remain separate and condense with ``s-wave'' symmetry: hence the
 insensitivity to scattering. The resulting condensate then
 automatically gives the spinons quasiparticle character, if,
 following Fisher at al, we suppose that the holons are boson-like
 rather than semions (thus returning to the original BZA
 hypothesis\cite{eleven}).  Admittedly, this hypothesis is speculative, but
it
 is strongly supported by experimental fact.

 \section{Conclusion and Questions}
 The existence of a quantum protectorate effects seems to me to be
 amply justified by the striking experimental anomalies I have
 listed: absence of phonon scattering, absence of pair breaking
 effects, the unusual phenomenon of the spin gap.
 These anomalies are   more
 strongly inescapable than many of the peculiarities often
 fastened on as the crucial key to understanding these very
 complex materials. Our version of the rather old phenomenon of
 charge-spin separation seems the most plausible source. We propose
 a new vision of CSS arising not from below, from the influence
 of a mysterious ``Quantum Critical Point'', but as being a
 universal high-energy trait of electron systems only renormalized
 away in low temperature and high dimensional systems.

 Many questions remain. Is the $Z_2$ symmetry which plays a key
 role in our view of CSS the same as the $Z_2$ which Senthil and
 Fisher find from a gauge theory analysis?
 It is very suggestive to say yes. If so, we presume that,
 contrary to their suggestion, even the superconducting state is
 one in which $Z_2$ symmetry is broken.

 The precise mechanism for the final superconducting transition is
 of course still in question since we have made no specification
 of the holon dynamics; I suggest that its $T_c$ is determined by
 the need to reduce the frustrated kinetic energy of the system,
but do not here propose an explicit mechanism.

\begin{figure}
\caption{
The Generalized Phase diagram. I-IV label the ``protected'' phases.
}
\caption{
Contrasting Self-Energies in a Fermi Liquid and a
non-Fermi liquid (from Ref.\  (2))
}
\end{figure}
\end{document}